# Spectroscopy of spontaneous spin noise as a probe of spin dynamics and magnetic resonance

S. A. Crooker[1], D. G. Rickel[1], A. V. Balatsky[2] & D. L. Smith[2]

[1]National High Magnetic Field Laboratory, Los Alamos, New Mexico 87545, USA

[2]Theory Division, Los Alamos National Laboratory, Los Alamos, New Mexico 87545, USA

**Not all noise in experimental measurements is unwelcome. Certain fundamental noise sources contain valuable information about the system itself -- a notable example being the inherent voltage fluctuations that exist across any resistor (Johnson noise), from which temperature may be determined[1,2]. In magnetic systems, fundamental noise can exist in the form of random spin fluctuations[3,4]. Felix Bloch noted in 1946 that statistical fluctuations of $N$ paramagnetic spins should generate measurable noise of order $\sqrt{N}$ spins, even in zero magnetic field[5,6]. Here we address precisely these same $\sqrt{N}$ spin fluctuations, using off-resonant Faraday rotation to passively[7,8] "listen" to the magnetization noise in an equilibrium ensemble of paramagnetic alkali atoms. These random fluctuations generate spontaneous spin coherences which precess and decay with the same characteristic energy and time scales as the macroscopic magnetization of an intentionally polarized or driven ensemble. Correlation spectra of the measured spin noise reveals g-factors, nuclear spin, isotope abundance ratios, hyperfine splittings, nuclear moments, and spin coherence lifetimes -- without having to excite, optically pump, or otherwise drive the system away from thermal equilibrium. These noise signatures scale inversely with interaction volume, suggesting routes towards non-perturbative, sourceless magnetic resonance of small systems.**

The fluctuation-dissipation theorem states that the response of a system to an external perturbation (*i.e.*, the susceptibility) can be described by the spectrum of fluctuations exhibited by the system in thermal equilibrium[9]. In magnetic systems, $\sqrt{N}$ fluctuations in an ensemble of *N* undriven nuclear spins was observed by Sleator and co-workers[6], validating Bloch's supposition. Fundamental magnetic fluctuations of thermal (and quantum) origin have since been identified in, *e.g.*, spin glasses[10], hard-disk magnetoresistive heads[11], and in the magnetic noise spectrum of antiferromagnetic particles[12]. Optical techniques have identified the presence of stochastic spin fluctuations in atomic systems[4,13-16], and notably, fluctuations corresponding to very few spins - and perhaps even single spins – are evidenced by ultrasensitive cantilevers[17] and scanning tunneling microscopes[18,19], respectively. In other disciplines, thermal noise in nanomechanical resonators[20] and recent correlation spectra of thermal acoustic vibrations suggest a means for 'sourceless ultrasonics'[21]. Here we investigate the detailed spectroscopy of spin noise to perform perturbation-free magnetic resonance. We study ground-state magnetization fluctuations in a classical ensemble of uncorrelated paramagnetic spins in thermal equilibrium, realized in atomic alkali vapors. Random spin fluctuations and their associated coherences reveal the complete magnetic structure of the atomic $^2S_{1/2}$ ground state, including hyperfine, Zeeman, and nuclear moment effects. Historically, this information is obtained with conventional magnetic resonance techniques (optical pumping and/or radio-frequency excitation)[22-24], which necessarily perturb the spin ensemble away from thermal equilibrium.

Figure 1a shows an experimental schematic. A Ti:sapphire ring laser (~8 GHz linewidth), detuned from any atomic absorption, is linearly polarized and focused through a cell containing a 40 mm long, temperature-controlled column of rubidium or potassium vapor having typical density of order $10^9$ atoms/mm$^3$. Random magnetization fluctuations along **z** impart small polarization (Faraday) rotation fluctuations $\delta\theta_F(t)$ on the laser, which are measured with a balanced photodiode bridge. Helmholtz coils provide a small transverse magnetic field B along **x**, about which all magnetization fluctuations $\delta M_z$ must precess. This precession, which corresponds to transverse spin coherences between Zeeman sublevels, shifts the spin fluctuation signal away from very low frequencies where environmental noise dominates. The detuned laser ensures a perturbation-free probe[7] of equilibrium spin noise, wherein the ~$10^9$ atoms within the laser volume are not optically pumped or excited in any way. The ensemble exhibits zero net magnetization ($\langle M_z(t) \rangle = 0$), with nominally equal population of spins within ground state sublevels.

Near the D1 or D2 transition of alkali atoms, the indices of refraction for right- and left-circularly polarized light are $n^\pm - 1 \cong (N_0 e^2 f / 4\pi m \nu \Delta)(1 \pm \beta \langle J_e \rangle)$, where $N_0$ is the density of atoms, $f$ and $\beta$ are the transition's oscillator strength and polarizability ($\beta$=2 and –1 for D1 and D2, respectively), $\nu$ is the laser frequency, and $\Delta = \nu - \nu_0$ is the laser detuning from transition center (assumed here to be much larger than pressure-broadened transition width). $\langle J_e \rangle \equiv \langle N_0^+ - N_0^- \rangle / 2N_0$ is the expectation value of the valence electron spin along **z**, where $N_0^\pm$ are the densities of atoms with ground-state spin projection parallel and antiparallel to the laser. The number of atoms within the laser beam of cross sectional area $A$ and over length $L$ is $N = N_0 A L$. Magnetization noise arises from statistical fluctuations in the quantity $N^+ - N^-$, which has amplitude $\sqrt{\langle (N^+ - N^-)^2 \rangle} = \sqrt{N} = \sqrt{N_0 A L}$. The induced Faraday rotation, $\theta_F = \pi \nu L (n^+ - n^-)/c$, therefore exhibits fluctuations

$$\langle \delta \theta_F^{\,2} \rangle^{1/2} = \frac{e^2 f \beta}{4mc\Delta} \sqrt{\frac{N_0 L}{A}}. \quad \text{(Eq.1)}$$

The spin correlation function, $S(t) = \langle M_z(0) M_z(t) \rangle$, has a Fourier transform $S(\omega)$ that is proportional to the measured power spectrum of $\delta\theta_F(t)$.

A typical noise spectrum from rubidium vapor is shown in Fig. 1b, taken with the laser detuned 25 GHz from the D1 transition. The sharp peaks at frequencies $\Omega$=869 and 1303 kHz are due to random spin fluctuations which are precessing in the small 1.85G transverse magnetic field, effectively generating spontaneous spin coherences between ground-state Zeeman sublevels. These coherences precess with effective g-factors $g_F = h\Omega/\mu_B B \sim 1/3$ and 1/2, which are the ground-state g-factors of the stable isotopes $^{85}$Rb and $^{87}$Rb ($h$ and $\mu_B$ are the Planck constant and Bohr magneton, respectively). Coupling of the nuclear spin $I$ to the $J=½$ valence electron splits the $^2S_{1/2}$ atomic ground state into two hyperfine $F$-levels with total spin $F = I \pm J$ and g-factor $|g_F| \cong g_J/(2I+1)$, where $g_J \cong 2$ is the free electron g-factor. Thus, the nuclear spin of $^{85}$Rb ($I=5/2$) and $^{87}$Rb ($I=3/2$) may be directly measured from spin fluctuations in thermal equilibrium. Noise spectra acquired near the D2 transition show similar peaks (inset, Fig. 1b), which move as expected with magnetic field. The 13 kHz measured width of these noise peaks indicates an effective transverse spin dephasing time ~100 μs, much less than the known Rb spin lifetime (~1s), due largely to the transit time of atoms across the ~100 μm laser diameter. The spectral density of the spin noise is small -- the $^{87}$Rb peak in Fig. 1b contributes only 3.1 nrad/$\sqrt{Hz}$ of Faraday rotation noise above the photon shot noise floor of 23 nrad/$\sqrt{Hz}$. Because spin noise arises, effectively, from $N$ uncorrelated precessing spins, the integrated area under the noise peaks should scale with $\sqrt{N}$. This is conveniently confirmed by noting that the integrated spin noise of the $^{85}$Rb and $^{87}$Rb peaks is 24.7 and 15.4 μV respectively, whose ratio - 1.60 - agrees quite well with the *square root* of the $^{85}$Rb:$^{87}$Rb natural abundance ratio ($\sqrt{72.2\%/27.8\%} = 1.61$).

That the off-resonant laser non-perturbatively probes spin fluctuations is evidenced in Fig. 2, where the integrated $^{85}$Rb spin noise is measured at large detuning from both the D1 and D2 transitions. The measured D2 absorption (optical depth) is also shown for reference. Appreciable spin noise exists out to $\Delta = \pm 100$ GHz, increasing at smaller detunings down to 15 GHz, at which point noticeable absorption occurs. For $|\Delta| > 20$ GHz, the measured spin noise scales as $\Delta^{-1}$, confirming that the off-resonant laser is sensitive to ground-state spin fluctuations through the dispersive index of refraction (as distinct from techniques in which laser noise is intentionally and resonantly absorbed by the system[25,26]).

The amplitude of fluctuations from $N$ uncorrelated spins scales as $\sqrt{N}$, as has been observed in atomic cesium[4,15] where, for quantum nondemolition measurements of atomic spin, such noise naturally imposes an obstacle to precision measurements. Here, $\sqrt{N}$ scaling is explicitly verified in Figs. 3a,b by temperature-tuning the Rb vapor density $N_0$. A power-law fit to the integrated spin noise with slope ½ shows excellent agreement. Thus, the ratio of spin noise *relative* to the signal one would measure in a magnetized sample ($\sim \sqrt{N}/N$) necessarily increases in systems with fewer spins. It is noteworthy, then, that the *absolute* noise increases when the size of the probe laser *shrinks* (Fig. 3c,d). At fixed density $N_0$, the effective interaction volume (and hence N) is reduced by shrinking the cross-sectional area $A$ of the laser while maintaining constant laser power. In this case, the $^{85}$Rb spin noise increases. This result may be viewed as a consequence of the $A^{-1/2}$ term in Eq. 1, or by considering that the Faraday rotation imposed on light passing through an intentionally-magnetized system is independent of beam area, so that the effective measurement sensitivity (rotation angle per unit polarized spin, $\theta_F/N$) is larger for narrower beams. Therefore, fluctuations of order $\sqrt{N}$ spins induce correspondingly more signal. Figure 3d shows the total spin noise versus beam area, with a power-law fit to $1/\sqrt{A}$ showing excellent agreement. These data suggest the utility of noise spectroscopy for passive probes of small systems, where the absolute amplitude of measured fluctuations actually increases when probe size is reduced, as long as measurement sensitivity increases correspondingly. In magnetometry this situation can be realized, for example, through the use of smaller Hall bar magnetometers (since the Hall voltage is independent of area, for fixed current), or as is often the case for magneto-optical measurements, through a tighter focus.

Finally, we show that fluctuation correlation spectra can reveal detailed information about complex magnetic ground states arising from, *e.g.*, nuclear magnetism and hyperfine interactions. Figure 4a shows the spin noise spectrum in $^{39}$K. With increasing magnetic field, the noise peak broadens and eventually splits into 4 resolvable spin coherences. This is the well-known quadratic Zeeman effect, which originates in the gradual decoupling of the electron and nuclear spin (the hyperfine interaction) by the applied magnetic field. Within each hyperfine level (see Fig. 4b), the 2F+1 Zeeman levels become unequally spaced, resulting in 2F distinct coherences between adjacent sublevels (Supplementary Figure). The energies of the Zeeman levels are given by the Breit-Rabi formula[24],

$$E_{F,m_F} = \frac{-\Delta_{hf}}{2(2I+1)} - g_I \mu_N B m_F \pm \frac{\Delta_{hf}}{2}\sqrt{1 + \frac{4m_F}{2I+1}x + x^2},$$

where $x = (g_J \mu_B + g_I \mu_N)B/\Delta_{hf}$, $\Delta_{hf}$ is the zero-field hyperfine splitting, $m_F$ is the magnetic quantum number, $g_I$ is the nuclear g-factor, $\mu_N$ is the nuclear magneton, and the $\pm$ refers to the upper and lower ($F = I \pm \frac{1}{2}$) hyperfine state. Without nuclear effects ($g_I = 0$), transitions within the two hyperfine manifolds are degenerate and $\Delta_{hf}$ can be measured via the separation between noise peaks, $\delta\Omega = 2\Omega_0^2/\Delta_{hf}$. When nuclear terms are included, transition energies within the upper and lower hyperfine manifolds are no longer exactly degenerate. This additional nuclear Zeeman splitting $\delta E_{nuclear}$ is resolved in the noise spectrum of $^{87}$Rb at 38 G, where all six allowed $\Delta F = 0, \Delta m_F = \pm 1$ magnetic resonance transitions are observed (Fig. 4c). The nuclear moment ($\mu_I$) and g-factor can be determined through $\delta E_{nuclear} = 2g_I \mu_N B = 2\mu_I B/I$. Lastly, spin fluctuations in thermal equilibrium also generate spontaneous spin coherence between *inter-hyperfine* Zeeman levels ($\Delta F = 1, \Delta m_F = \pm 1$), as shown in Fig. 4d for $^{39}$K. At low fields, these high-frequency noise coherences split away from the $^{39}$K hyperfine frequency ($\Delta_{hf}$ =461.7 MHz) with energy $h\Omega \cong \pm g_F \mu_B B$ and $\pm 3 g_F \mu_B B$, providing additional means of measuring $\Delta_{hf}$ from spin fluctuations alone.

We emphasize that these measurable ground-state spin coherences arise solely from random fluctuations while in thermal equilibrium, decidedly in contrast with normal methods for magnetic resonance. Nonetheless, the same detailed spectroscopic information is revealed, in accord with the fluctuation-dissipation theorem. The non-perturbative nature and inverse scaling of absolute noise with probe size portends favorably for local noise spectroscopy of small solid-state systems, where the planar geometry of many technologically-relevant structures is well-suited to high-spatial resolution studies. For example, in a semiconductor two-dimensional electron gas with electron density $10^{11}$ cm$^{-2}$, only ~1000 electrons are probed in a focused 1 micron laser spot. Thus, electron spin fluctuations (relative to the signal from a polarized system) are of order one part in $\sqrt{1000}$, as compared with only one part in $\sim \sqrt{10^9}$ in these studies. Other possibilities include spectroscopy of local magnetization fluctuations in next-generation hard-disk read/write heads[11,27], heterostructures for "spintronic"[28] applications and quantum information processing[29], or proposed noise probes of phase transitions and electron correlation in quantum Hall systems[30], and even single atoms[19].

**Methods:**

Polarization fluctuations $\delta\theta_F(t)$ imposed on the transmitted laser beam (due to magnetization fluctuations in the atomic vapor) are measured in a balanced photodiode bridge. The bridge consists of a polarization beamsplitter oriented at 45 degrees to the incident laser polarization, and $\theta_F$ is measured via the normalized intensity difference between the two orthogonal components at $\pm 45°$: $2\theta_F \cong (I_{+45} - I_{-45})/(I_{+45} + I_{-45})$. The photodiode difference current is converted to voltage (transimpedance gain=40V/mA, or ~20V/mWatt of unbalanced laser power at 790 nm) and measured in a spectrum analyzer. Above a few kilohertz, the measured noise floor arises primarily from photon shot noise, which contributes ~175 nV/$\sqrt{Hz}$ of spectrally flat (white) noise at a typical laser power of 200 µW. The photodiodes and amplifier contribute an additional 65 nV/$\sqrt{Hz}$ of uncorrelated white noise. All noise spectral densities are root-mean-square (rms) values, and spectra were typically signal-averaged for 10-20 minutes. The accuracy of measured hyperfine constants and g-factors was imposed by the $\pm 0.01$ gauss resolution of the Hall bar magnetic field sensor. Measurement of inter-hyperfine spin coherence (Fig. 4d) was performed with a higher-bandwidth, lower gain amplifier (~0.70 V/mA), and a typical laser power of 3.5 mW. Unless otherwise stated, measurements of rubidium (potassium) vapor were performed with 250 (125) Torr of nitrogen buffer gas, which broadens the linewidth of the D1 and D2 optical transitions and causes the motion of the alkali atoms to become diffusive (thereby increasing the average time the atoms spend in the laser beam and narrowing the magnetic resonance peaks.) The diameter of the laser beam in the vapor cell can be increased (decreased) by closing (opening) the aperture in front of the final focusing lens.

**Supplementary Information** accompanies the paper on **www.nature.com/nature**.

**Acknowledgements** We thank P. Littlewood, S. Gider, P. Crowell and P. Crooker for discussions. This work was supported by the Los Alamos LDRD program.

**Figure 1** Spontaneous spin noise in Rb or K vapor, probed via Faraday rotation. **a**, Experimental schematic. Ground-state stochastic spin fluctuations $\delta M_z(t)$ impart polarization fluctuations $\delta\theta_F(t)$ on the detuned probe laser. **b**, Measured spectrum of spin (Faraday rotation) noise from Rb vapor at T=369K and B=1.85 G, showing spontaneous spin coherence peaks from $^{85}$Rb and $^{87}$Rb. The data are given in units of root-mean-square (rms) spectral density of voltage fluctuations (nanovolts/$\sqrt{Hz}$) measured in the photodiode bridge, and also by the rms spectral density of Faraday rotation fluctuations (nanoradians/$\sqrt{Hz}$). The noise floor is determined primarily by photon shot noise. The laser is detuned $\Delta_{D1}$=25 GHz from the D1 transition ($5^2S_{1/2}$ to $5^2P_{1/2}$, ~794.8 nm), ensuring negligible absorption. Inset: The $^{85}$Rb and $^{87}$Rb spin noise peaks measured at $\Delta_{D2}$=20 GHz from the D2 transition ($5^2S_{1/2}$ to $5^2P_{3/2}$, ~780 nm). The magnetic field B is incremented in steps of 2.7 G, and T=369 K. Plots offset vertically for clarity.

**Figure 2** Total integrated spin noise from $^{85}$Rb as a function of laser detuning from the D1 (black points) and D2 (blue points) transitions. The laser power was 200 µW, T=359.1K, and the D1 (D2) transition was pressure broadened with 125 Torr (250 Torr) of $N_2$ buffer gas. Lines are guides to the eye. For reference, the dotted red line is the measured D2 optical depth (~2.5 at peak, or ~92% absorption).

**Figure 3** Spin noise from $^{85}$Rb versus spin density and beam cross-section. **a**, Spin fluctuation spectrum at five different spin densities, tuned via temperature, with fixed beam area (0.030 mm$^2$, or 1.2 mm$^3$ volume), 150 µW laser power, $\Delta_{D1}$=25 GHz detuning, and B=5.8G. **b**, The corresponding integrated spin noise vs. spin density, showing scaling with $\sqrt{N_0}$ (red line). **c**, *Increasing* absolute spin noise with *decreasing* cross-sectional beam area (*i.e.*, area within the 1/e$^2$ intensity envelope of the laser). The spin density is fixed (T=364K), with constant 145 µW laser power, $\Delta_{D1}$=25 GHz detuning, and B=5.8G. **d**, The corresponding integrated spin noise vs. beam area, showing scaling with $1/\sqrt{A}$ (red line).

**Figure 4** The ground-state Zeeman and hyperfine structure of $^{87}$Rb and $^{39}$K as revealed by stochastic spin coherences. **a**, Broadening and eventual breakup of the $^{39}$K spin noise peak into spectrally distinct Zeeman coherences due to the quadratic Zeeman effect. The magnetic field B=0.81, 2.88, 5.27, and 8.85 G, and the data are plotted relative to the center frequency $\Omega_0$ of the spin noise peak, as labeled. Nuclear Zeeman effects are not resolved, and the integrated spin noise is conserved (to within 5%), as expected. The laser is detuned $\Delta_{D1}$=-220 GHz from the $^{39}$K D1 transition (770 nm), and T=456K. **b**, Schematic diagram of the ground state hyperfine and Zeeman levels of $^{87}$Rb ($\Delta_{hf}$ =6835 MHz) and $^{39}$K ($\Delta_{hf}$ =461.7 MHz), which both have nuclear spin *I=3/2*. **c**, Ground-state spin fluctuation spectrum of $^{87}$Rb in B=38.1 G. Spin coherences between all allowed $\Delta F = 0$, $\Delta m_F = \pm 1$ Zeeman transitions are resolved, from which the hyperfine splitting and nuclear magnetic moment may be inferred. Transitions within the F=2 (F=1) ground-state hyperfine level are labeled in black (red). **d**, Spontaneous *inter*-hyperfine spin coherences ($\Delta F = 1$, $\Delta m_F = \pm 1$) in the noise spectrum of $^{39}$K at B=0.81, 2.88, and 5.27 G. The laser is detuned by $\Delta_{D1}$=-220 GHz, and T=456K.

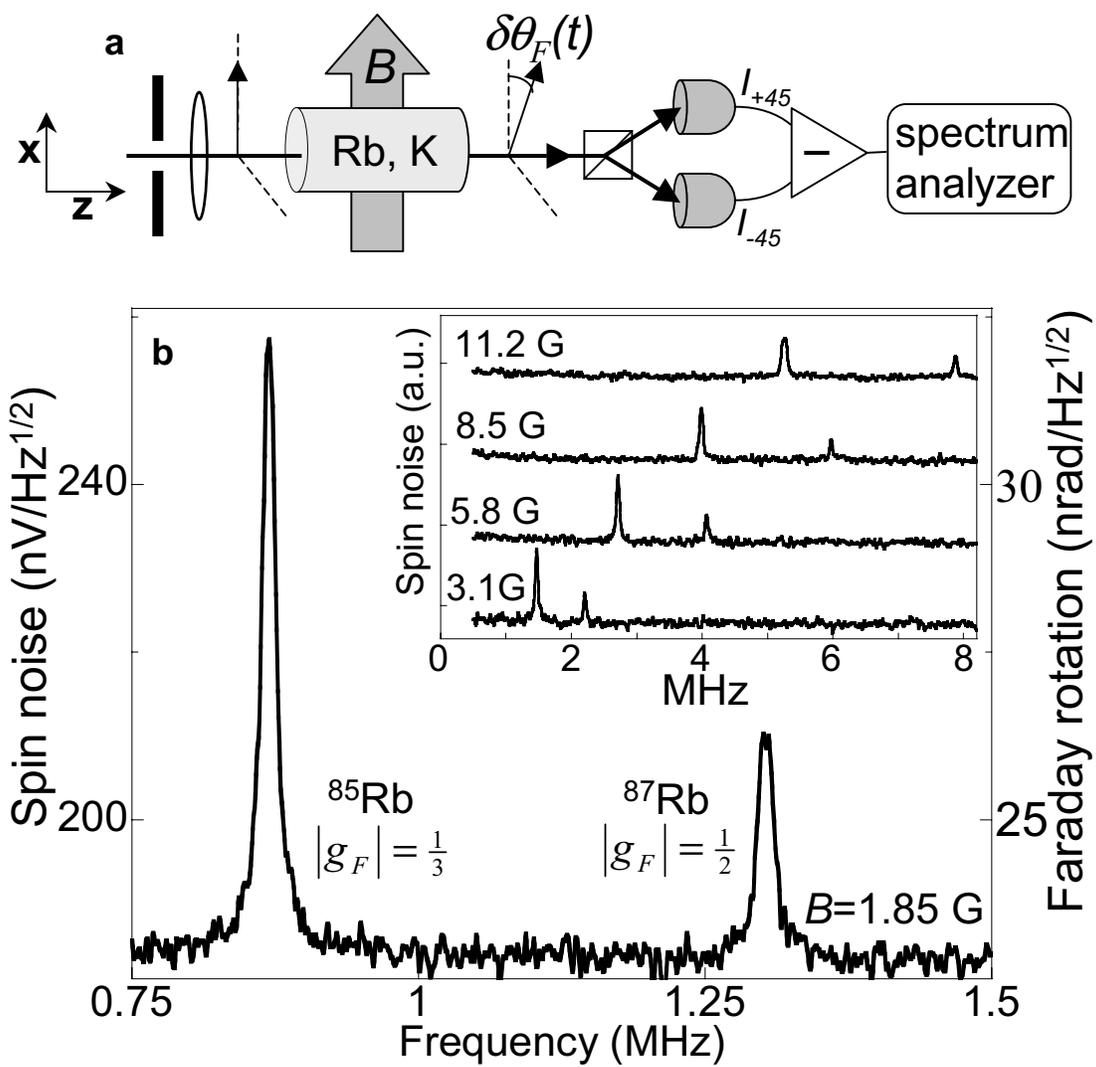

Figure 1
Crooker *et al*

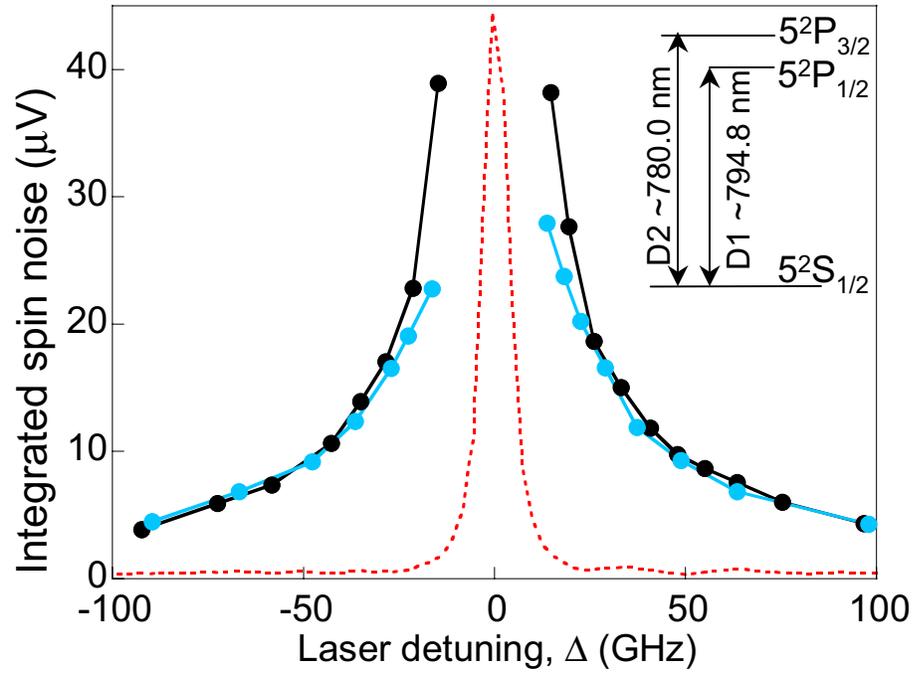

Figure 2
Crooker *et al*

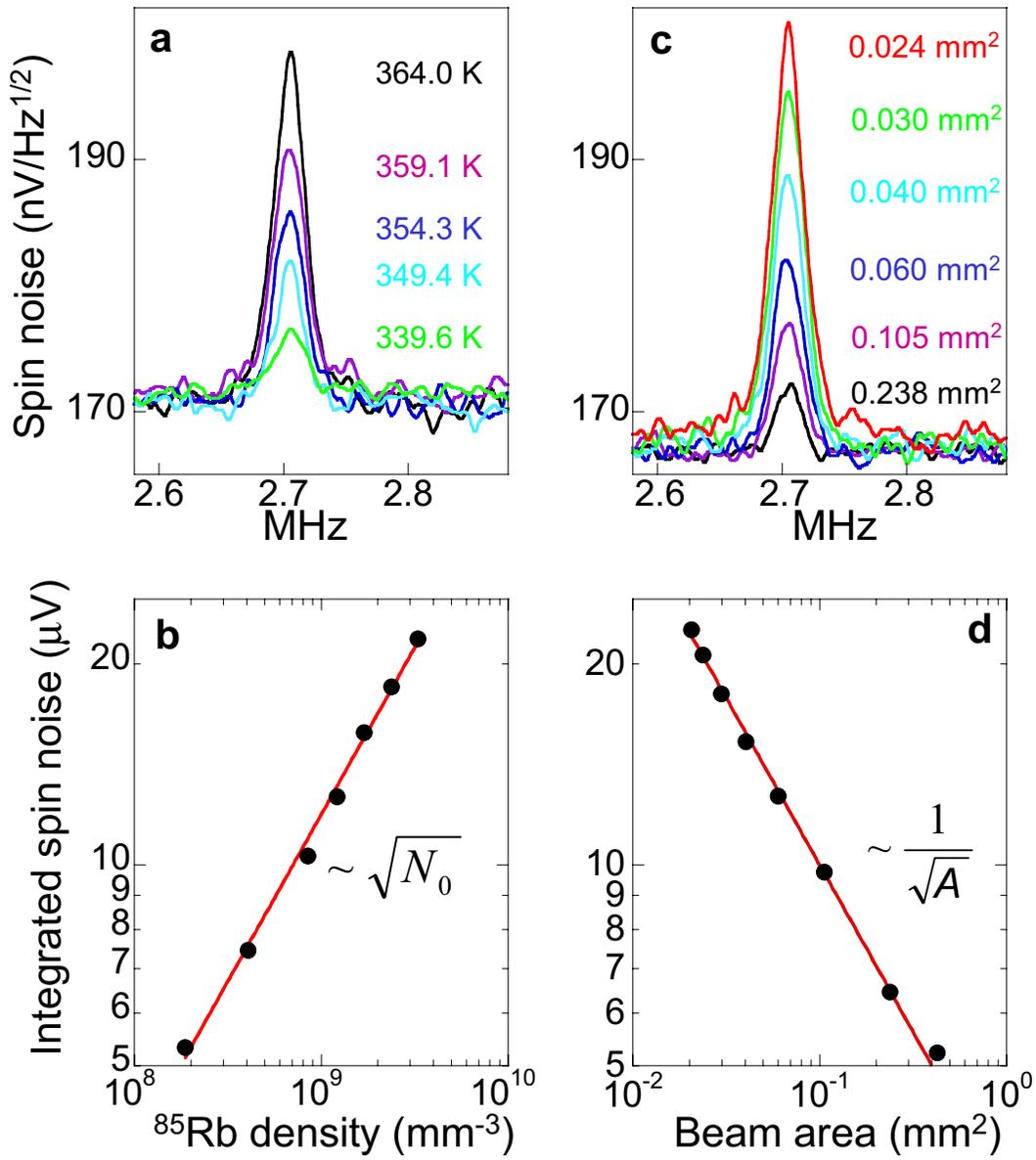

Figure 3
Crooker et al

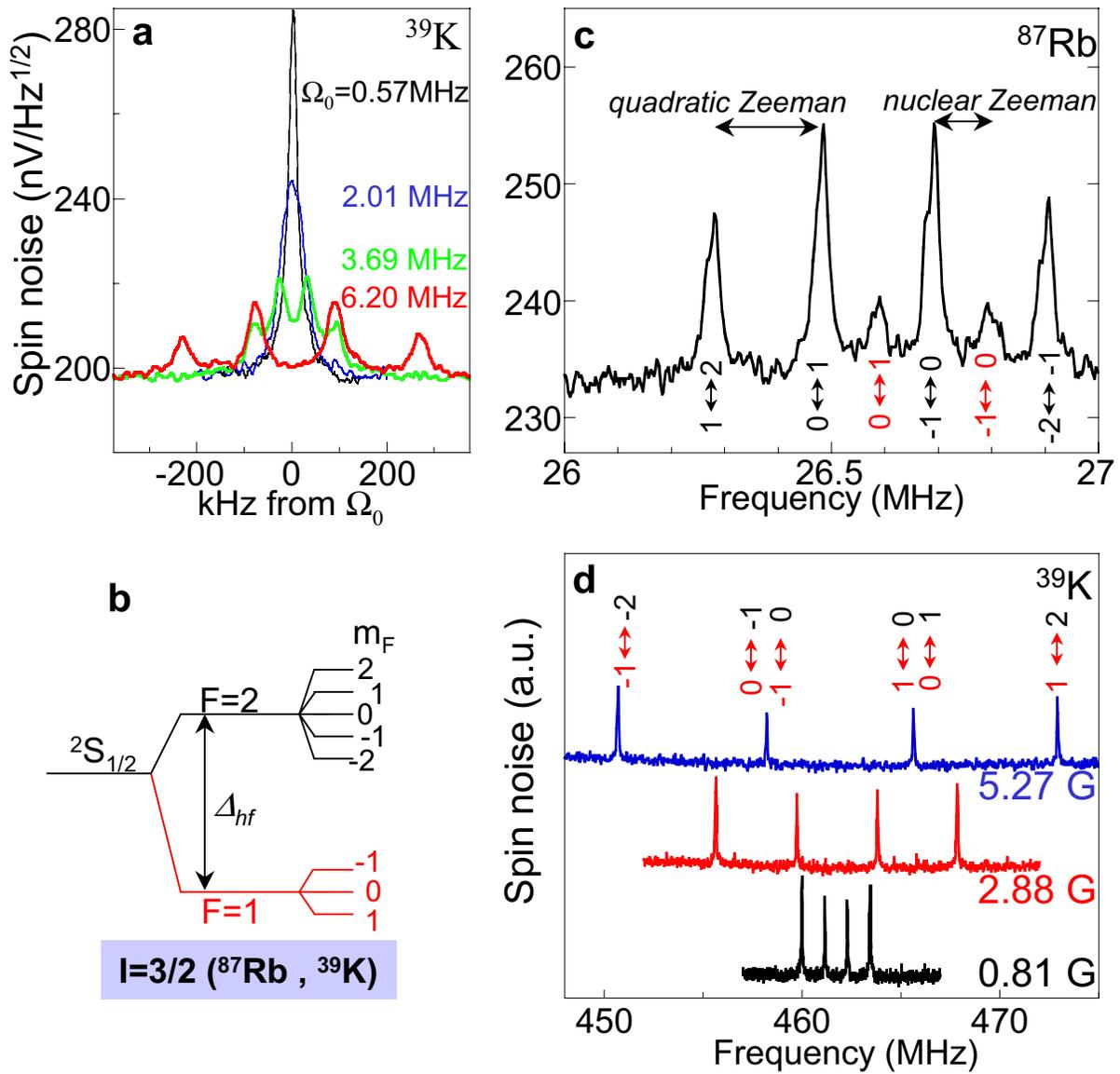

Figure 4
Crooker et al

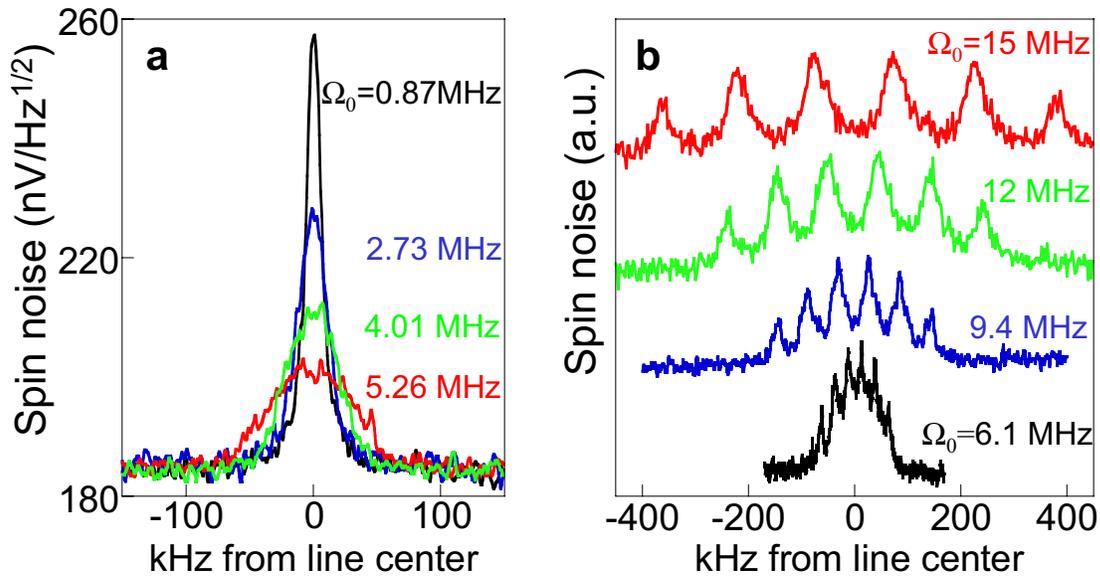

**Supplementary Figure** Evolution of spin noise from atoms with nuclear spin I=5/2 and ground-state hyperfine levels F=3 and F=2 ($^{85}$Rb). **a**, Broadening of the $^{85}$Rb spin noise with increasing magnetic field (B=1.85, 5.8, 8.5, and 11.2 G), at constant detuning $\Delta_{D1}$=25 GHz and laser power (175 µW). Data are plotted relative to the center frequency $\Omega_0$ of the noise peak, as labeled. The integrated spin noise is constant to within 5%. **b**, At higher magnetic fields (B=13.1, 20.1, 25.9, and 32.2 G), breakup of the $^{85}$Rb noise peak into 6 distinct noise coherences due to quadratic Zeeman effect. Nuclear Zeeman effects are not resolved (the 4 $\Delta m_F$=1 coherences within the lower F=2 hyperfine level are merged with the middle 4 coherences from the upper F=3 hyperfine level). Detuning and laser power vary between traces, and plots are offset for clarity.